\documentclass[a4paper,12pt]{article}
\usepackage{mathptm}
\usepackage{setspace}
\usepackage{color}
\usepackage{graphicx}
\usepackage{pdfcolmk}   
\usepackage{multicol}

\def\url#1{\textcolor{blue}{\underline{#1}}}

\begin{document}


\thispagestyle{empty}

\begin{centering}

{\it Biophysical Journal}, in press (2007) \\ \ \\ 

{\LARGE \ \\ \ \\ A modified cable formalism for modeling \\ neuronal
membranes at high frequencies}

{\large \ \\ Claude B\'edard and Alain
Destexhe\footnote{Corresponding author -- Destexhe@iaf.cnrs-gif.fr}
\\ \ \\ Integrative and Computational Neuroscience Unit (UNIC), \\
CNRS, Gif-sur-Yvette, France \\ \ \\ \today \\ \ \\ }

\end{centering}


\begin{abstract}

Intracellular recordings of cortical neurons {\it in vivo} display
intense subthreshold membrane potential (V$_m$) activity. The power
spectral density (PSD) of the V$_m$ displays a power-law structure at
high frequencies ($>$50~Hz) with a slope of about -2.5. This type of
frequency scaling cannot be accounted for by traditional models, as
either single-compartment models or models based on reconstructed
cell morphologies display a frequency scaling with a slope close to
-4. This slope is due to the fact that the membrane resistance is
``short-circuited'' by the capacitance for high frequencies, a
situation which may not be realistic.  Here, we integrate non-ideal
capacitors in cable equations to reflect the fact that the
capacitance cannot be charged instantaneously. We show that the
resulting ``non-ideal'' cable model can be solved analytically using
Fourier transforms. Numerical simulations using a ball-and-stick
model yield membrane potential activity with similar frequency
scaling as in the experiments.  We also discuss the consequences of
using non-ideal capacitors on other cellular properties such as the
transmission of high frequencies, which is boosted in non-ideal
cables, or voltage attenuation in dendrites. These results suggest
that cable equations based on non-ideal capacitors should be used to
capture the behavior of neuronal membranes at high frequencies.

\end{abstract}


\clearpage

\section{Introduction}

One of the greatest achievements of computational neuroscience has
been the development of cable theory (reviewed
in~\cite{Rall95,Johnston95}), and which can explain many of the
passive properties of neurons, including how dendritic events are
filtered by the cable structure of dendrites.  Cable theory describes
the space and time propagation of the membrane potential by partial
differential equations.  Such a formalism constitutes the basis of
nearly all of today's computational models of dendrites, and is
simulated by several publically-available and widely-used simulation
environments (reviewed in~\cite{Rev2006}).

Some experimental observations, however, may suggest that the
standard cable formalism may not be adequate to simulate the fine
details of dendritic filtering.  One of these observations is the
fact that the power spectral density (PSD) of synaptic background
activity or channel noise does not match that predicted from cable
theory~\cite{Dest2003,Diba2004,Jacobson2005,Rud2005}.  The PSD scales
approximately as 1/f$^{\alpha}$ with an exponent $\alpha = 2.5$, both
for channel noise and background activity (Fig.~\ref{spectra}A-B),
whereas cable theory would predict scaling with an exponent $\alpha =
4$ or $\alpha = 5$ for synaptic inputs distributed in
dendrites~\cite{Diba2004,Pow2003}; see also Appendix~1), or $\alpha$
= 3.2 to 3.4 when inputs are distributed in soma and dendrites (see
Fig.~\ref{spectra}C-D). In other words, these data suggest that
frequencies are filtered by dendritic structures in a way different
from that predicted by traditional cable equations.

\ \\ \centerline{------------------------ Figure~\ref{spectra} here
------------------------}

One possible origin of such a mismatch could be due to the fact that
the permittivity of the membrane is frequency
dependent~\cite{Cole-Cole1941,White1970}.  However, capacitance
measurements in bilipid membranes shows negligible variations around
100~Hz (see Fig.~5 in~\cite{White1970}), suggesting that the
Cole-Cole model may not be the correct explanation for this range of
frequencies.  It could also be that distortions of the frequency
dependence arise from the complex three-dimensional morphology of the
neuronal membrane~\cite{Eisenberg1980}.  However, NEURON simulations
of the standard cable model using three-dimensional morphologies of
cortical pyramidal neurons give frequency scaling with an exponent
$\alpha$ $> 3$ (Fig.~\ref{spectra}C-D), suggesting that this is not a
satisfactory explanation either.

None of the previous models take into account the fact that the
surface of neuronal membranes is a complex arrangement, not only of
phospholipids, but also of a wide diversity of surface
molecules~\cite{Alberts2002}.  This complex surface may be
responsible for additional resistive phenomena not taken into account
in previous approaches.  In other words, the neuronal membrane may
not be an ``ideal'' capacitor, as commonly assumed in the standard
cable formalism.  In the present paper, we explore this hypothesis as
an alternative mechanism to explain the observed frequency scaling
and consider neuronal membranes as ``non-ideal'' capacitors.  We show
that cable equations can be extended by including a non-ideal
resistive component (Maxwell-Wagner time) in the capacitor
representing the membrane, and that the non-ideal cable model
reproduces the observed frequency scaling.  We also show consequences
of this extension to cable equations in voltage attenuation and
synaptic summation.  Our aim is to provide an extended cable
formalism which is more adapted to capture membrane potential
dynamics and dendritic filtering at high frequencies.  Some of these
results have appeared in a conference abstract~\cite{SFN2007}.


\section{Materials and Methods}

The standard and non-ideal cable equations were either solved
analytically (see Results) or simulated using custom-made programs
written in MATLAB.  A ``ball-and-stick'' model consisting of a
soma connected to a dendritic cylinder of length $l_d$ was simulated
(see Results for details).  Away from the current source, we have
the following equations (in Fourier space):
\begin{eqnarray}
   \lambda^2~\frac{\partial^2 V_m(x,\omega)}{\partial x^2}
      = \kappa_{ext}^2(\omega)V_m(x,\omega)
                                              \label{ext-cable1} \\
  \kappa_{ext}^2(\omega) = 1+i \frac{\omega\tau_m}{1 + i\omega\tau_M}    \nonumber
\end{eqnarray}
where $\lambda = \sqrt{{r_m}/{r_i}}$ is the electrotonic constant
that characterizes the cable, $\tau_m$ is the membrane time constant,
and $\tau_M$ is the Maxwell-Wagner time constant ($\tau_M = 0$
corresponds to the standard cable equations; see Results).

The ``source'' synaptic current consisted in a random synaptic
bombardment of Poisson-distributed synaptic events.  Each synaptic
event consisted of an instantaneously rising current followed by
exponential decay, and were summated linearly:
\begin{equation}
  I_S \ = \ A \ \sum_i \ H(t-t_i) \ \exp[-(t-t_i)/\tau_S] ~ ,
\end{equation}
where $I_S$ stands for the source current, $H(t)$ is the Heaviside
function, and $t_i$ are the times of each synaptic event
(Poisson-distributed with mean rate of 100~Hz).  The decay time
constant was $\tau_S$ = 10~ms and the amplitude of the current was
$A$ = 1~nA.

The source current was inserted at different positions $l_s$ in the
dendrite (see Results).  The voltage at the soma was obtained by
solving either standard or non-ideal versions of cable equations (see
Results and Appendix~2). The power spectral density (PSD) was
calculated from the somatic membrane potential using the fast Fourier
transform algorithms present in MATLAB (Signal Analysis toolbox).
The same algorithm was also used to calculate the PSD from
experimental data.

The experimental PSD of V$_m$ activity shown here were obtained from
intracellular recordings of cat parietal cortex neurons {\it in vivo}
and were taken from previous publications~\cite{Dest2003,Rud2005},
where all methodological details were given.  No filter was used
during digitization of the data, except for a low-pass filter with
5~kHz cutoff frequency during acquisition (sampling frequency of
10~kHz).  Thus, the PSD is expected to reflect the real power
spectral content of recorded V$_m$ up to frequencies of 4-5~kHz.

Some simulations (Fig.~\ref{spectra}C-D) were realized using
morphologically reconstructed neurons from cat cortex obtained from
two previous studies~\cite{Contreras97,Douglas91}, where all
biological details were given.  The three-dimensional morphology of
the reconstructed neurons was incorporated into the NEURON simulation
environment, which enables the simulation of the traditional cable
equations using a three-dimensional structure with a controlled level
of spatial accuracy~\cite{Hines97}.  Simulations of up to 3500
compartments were used.  {\it In vivo}--like activity was simulated
using a previously published model of synaptic bombardment at
excitatory and inhibitory synapses~\cite{Destexhe99} (see this paper
fo details about the numerical simulations).


\section{Results}

We start by deriving the non-ideal cable model, then investigate its
general properties by evaluating the PSD of somatic voltage, as well
as voltage attenuation.

\subsection{Derivation of non-ideal cable equations}

\subsubsection{The membrane as a non-ideal capacitor}

In electrostatics, if an electric field is applied to a closed
conductive surface, electric charges migrate until they reach
equilibrium (when the field tangential to the surface is zero). In
particular, the electric resistivity of the membrane imposes a given
velocity to charge movement, which dissipates calorific energy
similar to a friction phenomenon.  This calorific dissipation is
usually neglected, which amounts to consider an instantaneous charge
re-arrangement following changes in electric field.

However, in reality this calorific dissipation may have significant
consequences, and this phenomenon is well known for
capacitors~\cite{Bowick82}.  A ``non-ideal'' capacitor dissipates
calorific energy when the electric potential varies, and capacitors
are usually conceived such as to minimize this phenomenon and realize
the well-known ideal relation $i = C\frac{dV}{dt}$. A ``non-ideal"
linear capacitor can be represented as an arrangement of resistances,
inductance and capacitance (see Fig.~\ref{capacitor}A).  A linear
approximation, which is usually sufficient for most purposes.  In
particular, this approximation is valid when the effects of
electrostriction are negligible~\cite{White1970,Alvarez1978}.  This
is the case when the propagated signals are of small amplitude
(millivolts), because $C(V)=C(0) \ (1+a V^2)$, with typically $a$ =
0.02~$V^{-2}$~\cite{Alvarez1978}. In such cases, the membrane
capacitance can be represented by a resistance and a capacitance in
series~\cite{Raghuram1990} (see Fig.~\ref{capacitor}B).  The
resistance represents here the loss of calorific energy associated
with charge movement.  In standard cable equations, such a resistance
is not present (see Fig.~\ref{capacitor}C).

Thus, we use a more realistic capacitor modeled by taking into
account an additional resistance ($R_{sc}$), which accounts for the
calorific loss and the consequent finite-velocity of charge
rearrangement.  This R-C circuit will be characterized by a
relaxation time $\tau_M = R_{sc} C$, called ``Maxwell time" or
``Maxwell-Wagner time"~\cite{Raju2003,Bed2006}.  The Maxwell time
corresponds to the characteristic displacement time of the charges in
the capacitor.  Thus, such a non-ideal capacitor cannot be charged
instantaneously; the resistance $R_{sc}$ imposes a minimal charging
time due to finite charge velocities.

\ \\ \centerline{------------------------ Figure~\ref{capacitor} here
------------------------}

This phenomenon of finite charge velocity is particularly relevant to
biological membranes, which are capacitors in which charges are also
subject to rearrangements.  In the following, we attempt to include
this contribution to membrane capacitors by including Maxwell-Wagner
time to cable equations and determine its consequences.

\subsubsection{Non-ideal cable equations}

We extend cable equations by including a finite charge velocity
(or equivalently, a minimal charging time) to membrane capacitors.
We start by Ohm's law, according to which the axial current $i_i$
in a cylindric cable can be written as:
\begin{equation}
    i_i = \sigma\vec{E}= -\frac{1}{r_i}\frac{\partial V_m}{\partial x}~.
\end{equation}

We also have, for the membrane current $i_m$:
\begin{equation}
    i_m = -\frac{(i_i(x +\Delta x)-i_i(x))}{\Delta x}\approx -\frac{\partial i_i}{\partial x}
    ~ ,
\end{equation}
and we can write
\begin{equation}\label{im}
    i_m=\frac{V_m}{r_m}+
    \int_{-\infty}^{\infty}\frac{\partial c_m(t-t')}{\partial t}V_c(t')dt'
\end{equation}
where $c_m(t)$ is the inverse complex Fourier transform of the
capacitance $c_m(\omega )$.  $c_m(t ) = c_m \delta (t)$ if the
capacitance does not depend on the frequency.

Integrating Maxwell-Wagner phenomena, we have:
\begin{eqnarray*}
    V_m & = & V_c + r_{sc}\int_{-\infty}^{\infty}\frac{\partial c_m(t-t')}{\partial
t}V_c(t')dt'
\end{eqnarray*}
Thus, we obtain the following {\it non-ideal cable equations}:
\begin{eqnarray}
   \lambda^2~\frac{\partial^2 V_m}{\partial x^2}
      = V_m + r_{m} \int_{-\infty}^{\infty}\frac{\partial c_m(t-t')}{\partial t}V_c(t')dt'
                                              \label{ext-cable} \\
   V_m =r_{sc}~\int_{-\infty}^{\infty}\frac{\partial c_m(t-t')}{\partial t}V_c(t')dt' + V_c
                                         ~ ,     \nonumber
\end{eqnarray}
where $\lambda = \sqrt{{r_m}/{r_i}}$ is the electrotonic constant
that characterizes the cable.

\subsubsection{General solution of non-ideal cable equations}

The non-ideal cable equations (Eqs.~\ref{ext-cable}) are a linear
system with constant coefficients which can be solved by using
Complex Fourier Transforms:
\begin{eqnarray*}
    v_m(x,\omega) & = & \int_{-\infty}^{\infty} V_m(x,t)~e^{i\omega t}~dt \\
    v_c(x,\omega) & = & \int_{-\infty}^{\infty} V_c(x,t)~e^{i\omega
    t}~dt \\
    c_m(\omega) & = & \int_{-\infty}^{\infty} c_m(t)~e^{i\omega t}~dt
\end{eqnarray*}

We obtain the following expression:
\begin{equation}\label{eqaA}
   \lambda^2~\frac{d^2 v_m(x,\omega)}{d x^2} = \kappa_{ext}^{2} v_m(x,\omega)
\end{equation}
with
\begin{equation}\label{kappa}
 \kappa_{ext}^2 = {1 + i\frac{\omega\tau_m}{1 +i\omega\tau_M}} ~ ,
\end{equation}
where $\tau_m(\omega) = r_m c_m(\omega)$ and $\tau_M(\omega) =
r_{sc} c_m(\omega)$ are the parameters that characterize the
cable.

The general solution of Eq.~\ref{eqaA} is given by
\begin{equation}
  v_m(x,\omega)= A(\omega)exp(\frac{\kappa_{ext} (l_s -x)}{\lambda})
  + B(\omega) exp(-\frac{\kappa_{ext} (l_s -x)}{\lambda})
  \label{gen}
\end{equation}
where $l_s$ is the position of the current source in the dendrite.

This solution is similar to that of traditional cable equation,
with the only difference in the value of $\kappa$.  In cable
equations, this value is given by
\begin{equation}
 \kappa_{s}^2 = {1 + i\omega\tau_m} ~ .
 \label{std}
\end{equation}
In particular, for null frequency, the two cable formalisms are
equivalent
\begin{equation}
    \kappa_{ext}(0)=\kappa_{s}(0)=1~~,
\end{equation}
whereas they will predict different behavior for $\omega>0$.

In the following, we will consider that the capacitance is
independent of frequency, $c_m (\omega) = cst$, as also assumed in
the standard cable model~\cite{Rall95,Johnston95}.

Figure~\ref{kappafig} compares the values of $\kappa$ between the
two cable formalisms (with $c_m (\omega)$ = cst).  The difference
depends on the relative values of $\tau_M$ and $\tau_m$: for
$\tau_M << \tau_m$, the two formalisms are very similar, but
differ when $\tau_M$ is larger, in particular for high
frequencies.  Thus, the critical parameter is $\tau_M$, which
determines the saturation of the value of $\kappa$.

\ \\ \centerline{------------------------ Figure~\ref{kappafig} here
------------------------}

\subsection{Voltage attenuation vs.\ distance and frequency}
\label{attensec}

To compare the properties of the non-ideal cable model compared to
the standard cable model, we evaluated the properties of voltage
attenuation in a large dendritic branch.  We have chosen a cable of
$l_d =500~\mu m$ and diameter of 2~$\mu$m, with a current source
situated at one end of the cable ($x=l_s=0$) and connected to an
infinite impedance at the other end ($x=l_d$; ``sealed end'').  In
these conditions, we can determine the law of voltage attenuation
with distance, using complex Fourier analysis.

As we have seen above, the main difference between the standard and
non-ideal cable models lies in the expression for $\kappa$ (see
Eqs.~\ref{kappa} and \ref{std}).  In a finite cable of constant
diameter, the steady-state voltage attenuation profile is given by
the relation:
\begin{equation}\label{eqD}
   V_{m}(x,\omega) = A(\omega)~exp(-\frac{\kappa}{\lambda}x)
   +B(\omega)~exp(\frac{\kappa}{\lambda} x).
\end{equation}
for $x > 0$.  To evaluate the functions $A(\omega)$ and
$B(\omega)$, we apply the limit conditions of the dendrite.  At $x
=0$, we have a current source $i_s =1=i_d$, and at $x =l_d$ we
have $i_d =0$ (``sealed end'').  The expressions for $A$ and $B$
are then given by Eqs.~\ref{B5.8} and \ref{B5.9}, respectively
(see Appendix~2).

This relation is plotted in Fig.~\ref{atten} for two values of the
membrane time constant $\tau_m$ of 5~ms and 20~ms, which correspond
to two different conductance states of the membrane (the
corresponding electrotonic constant is $\lambda$ = 353.5~$\mu$m and
707.1~$\mu$m, respectively).  The voltage attenuation is in general
steeper for the non-ideal cable model, which effect is particularly
apparent for frequencies of the order of 0-50~Hz. However, this
effect reverses between 50 and 100~Hz, in which case the non-ideal
cable model shows a less steep voltage attenuation profile compared
to the standard cable model (see 50 and 100~Hz in Fig.~\ref{atten}).

\ \\ \centerline{------------------------ Figure~\ref{atten} here
------------------------}

\subsection{Power spectra of voltage noise predicted by non-ideal
cable equations} \label{vmPSD}

We now calculate the PSD of the voltage noise predicted by non-ideal
cable equations.  We consider a ``ball-and-stick'' model consisting
of a soma and a dendritic segment of variable length
(Fig.~\ref{ball-stick}A).  The source consists of a sum of
exponentially-decaying currents (see Materials and Methods), which
represent the synaptic current resulting from many synapses releasing
randomly, as shown in Fig.~\ref{ball-stick}B.  The source has a PSD
which scales as $1/f^\alpha$ with an exponent $\alpha=2$ at high
frequencies (Fig.~\ref{ball-stick}C).

\ \\ \centerline{------------------------ Figure~\ref{ball-stick} here
------------------------}

To investigate the PSD of the somatic voltage in the ball-and-stick
model, we first examine the PSD following a single source consisting
of summated exponential synaptic currents.  The standard cable model
predicts that such a source localized on a dendritic branch
(ball-and-stick model with $l_d$ = 500~$\mu$m and $\lambda \simeq
400~\mu$m) gives a V$_m$ PSD scaling approximately as $1/f^\alpha$
with an exponent $\alpha\simeq 4$, which corresponds to a somatic
impedance much larger than that of the dendrite (soma radius of
7.5~$\mu$m; see Appendix~1), which would correspond to most central
neurons for which the soma represents a minor proportion of the
membrane.  The V$_m$ PSD for the standard cable model with uniformly
distributed exponential synaptic currents is illustrated in
Fig.~\ref{PSDscaling} (continuous curve), and shows a frequency
scaling with an exponent $\alpha \simeq 4$.

In contrast, the non-ideal cable model gives different scaling
properties of the PSD, according to the value of $\tau_M$
(Fig.~\ref{PSDscaling}, dotted and dashed lines).  The power for high
frequencies ($>$50~Hz) is much larger in the non-ideal cable model
compared to the standard model, which shows that non-ideal cables
have enhanced signal propagation for high frequencies.  The V$_m$ PSD
for the non-ideal cable model with uniformly distributed exponential
synaptic currents is illustrated in Fig.~\ref{PSDscaling} (dashed
curve), and shows a frequency scaling with an exponent $2 < \alpha
\leq 4$ for $\tau_m \geq \tau_M \geq 0$, respectively ($\alpha \simeq
2$ when $\tau_m = \tau_M$, but it can be shown that $\alpha = 2$ only
if $\tau_M \rightarrow \infty$).

\ \\ \centerline{------------------------ Figure~\ref{PSDscaling} here
------------------------}

We next investigated the influence of the localization of the current
source in the dendrite.  Figure~\ref{sum}A shows the PSD obtained at
the soma of the ball-and-stick model when the current source was
placed at different positions in the dendrite.  The position affects
the amplitude of the PSD, and the frequency-scaling of the PSD is
affected by the position.  The scaling exponents obtained are of
$\alpha$ = 4.1416 for 250~$\mu$m and 5.3653 for 450~$\mu$m for the
standard model, and $\alpha$ = 2.5311 for 250~$\mu$m and 2.8354 for
450~$\mu$m for the non-ideal cable model.  The PSD obtained when
simulating a ``distributed'' synaptic bombardment in the dendrite
(Figure~\ref{sum}B) also displays the same frequency-scaling.
Similar results were also obtained by varying the parameters $\tau_m$
and $\tau_M$ (not shown), suggesting that the properties of frequency
scaling, as shown in Fig.~\ref{PSDscaling}, are generic.

\ \\ \centerline{------------------------ Figure~\ref{sum} here
------------------------}

To evaluate the optimal value of $\tau_M$ (for this particular model
with $\tau_m = 5~ms$), we fitted the PSD of the model to that of
experiments.  To perform this fit, we used a frequency range of 100
to 400~Hz, which was chosen such that it is not affected by
instrumental noise ($<$700~Hz) and such that the frequency band
considered belongs to the power-law scaling region of the spectra
($>$80~Hz).  The result of this fitting is shown in
Fig.~\ref{comparaison}.  The scaling exponent obtained are of
$\alpha$ = 3.6533 for the standard cable model, and of $\alpha$ =
2.3306 for the non-ideal cable model, for an optimal value of $\tau_M
= 0.3~\tau_m$.  This suggests that the calorific dissipation caused
by the resistivity of the membrane to charge movement is of the order
of 30\% of that caused by the flow of ions through ion channels.
This estimate is of course specific to the model used, but variations
of this model ($l_d$, diameter, number of dendrites, for a uniform
$\tau_m$ over the whole neuronal surface) showed little variation
around this value (not shown).

\ \\ \centerline{------------------------ Figure~\ref{comparaison} here
------------------------}

This value gives a cutoff frequency ($1/\tau_M$) around 105~Hz.  Above
this cutoff frequency, the membrane becomes more resistive than
capacitive because the energy loss due to calorific dissipation
becomes larger than the energy necessary for charge displacement.
This is very different than an ideal capacitor, in which the energy
from the current source would exclusively serve to charge
displacement. In Fig.~\ref{kappafig}, one can see that the value of
$\kappa$ for the non-ideal model departs from that of the standard
cable model around this cutoff frequency.

Thus, from the above figures, and especially Fig.~\ref{PSDscaling},
it is apparent that the non-ideal cable model has more transmitted
power compared to the standard cable model at high frequencies
($>>$100~Hz).  This increased transmission of high frequencies is
also visible by superimposing the V$_m$ activities of the standard
and non-ideal model (Fig.~\ref{Superpos}).  Such an increased
transmission at high frequencies can be explained by the fact that in
the standard cable model, the term $1/i \omega c_m$ tends to zero
when $\omega$ tends to infinity, such that for high frequencies $r_m$
is short-circuited by the capacitance of the membrane.  In the
non-ideal cable model, such a short-circuit does not occur, even at
frequencies much larger than the cut-off frequency.  This results in
a very different behavior at high frequencies, and a less pronounced
frequency fall-off in the non-ideal cable PSD.  Displacing charges by
capacitive effect takes energy, and this energy diminishes with
increasing frequencies in the non-ideal cable, which enables more
energy transfer between remote ion channels in dendrites (synapses
for example) and the soma at high frequencies.  This is also
consistent with the fact that the non-ideal cable equations display
less voltage attenuation (see Section~\ref{attensec}).

\ \\ \centerline{------------------------ Figure~\ref{Superpos}
here ------------------------}


\section{Discussion}

In the present paper, we have proposed an extension to the classic
cable theory to account for the behavior of neuronal membranes at
high frequencies.  Experimental observations indicate that the PSD of
the V$_m$ does not match that predicted from cable theory, in
particular for the frequency-scaling at high
frequencies~\cite{Dest2003,Diba2004,Jacobson2005,Rud2005}.  The
modification to cable equations consists in incorporating a
``non-ideal'' membrane capacitance by taking into account the
calorific dissipation due to charge displacement, which is usually
neglected.  We have shown that this ``non-ideal'' cable formalism can
account for the frequency scaling of the PSD observed experimentally
for high frequencies (Fig.~\ref{comparaison}).

In experiments with channel noise or synaptic noise, the $V_m$ PSD
scales as $1/f^\alpha$ with an exponent $\alpha$ around
2.5~\cite{Dest2003,Diba2004,Jacobson2005,Rud2005}.  The standard
cable model predicts that the somatic V$_m$ should scale with an
exponent $\alpha$ comprised between 3 and 4~\cite{Diba2004}, when the
source is located in the soma.  However, we have shown here that the
frequency scaling of the V$_m$ PSD depends on the location of the
source, and that the exponent $\alpha$ is equal or larger when
current sources are located in dendrites (see Fig.~\ref{sum} and
Appendix~1).  Thus, the standard cable model cannot account for
exponents lower than $\alpha$ = 3.  On the other hand, taking into
account non-ideal capacitances may lead to scaling exponents down to
$\alpha=2$, depending on the magnitude of the dissipation in the
non-ideal capacitance (as quantified by the value of the
Maxwell-Wagner time $\tau_M$; see Fig.~\ref{PSDscaling}).  In the
case $\tau_M$ is non-uniform, then one may have larger differences of
frequency scaling between somatic and dendritic current sources (not
shown).

In the non-ideal model, the calorific dissipation originates mostly
from the resistance of the membrane to lateral ion displacement. 
This ``tangential'' resistance is not yet characterized
experimentally and is equivalent to the resistance involved in the
non-instantaneous character of membrane polarization~\cite{Bed2006}. 
Several arguments indicate that this resistance may be substantial. 
First, the membrane surface contains various molecules such as sugars
and various macromolecules, in addition to
phospholipids~\cite{Alberts2002}.  Thus, lateral ion movement is
likely to be affected by collisions or tortuosity imposed by these
molecules.  Second, the phospholipids themselves contain local
dipoles at their polar end, which is likely to cause local
electrostatic interactions which may influence the lateral movement
of ions.  Indeed, the fitting to experimental data using the
non-ideal cable model predicts a value for $\tau_M$ which is a
significant fraction ($\sim$30\%) of the membrane time constant.

The complex three-dimensional membrane morphology could have
consequences on frequency-de\-pendent properties even with
traditional cable theory~\cite{Eisenberg1980}.  We tested this
possibility by simulating detailed three-dimensional morphological
models of cortical pyramidal neurons and failed to reproduce the
frequency scaling of the V$_m$ activity {\it in vivo} (see
Fig.~\ref{spectra}).  Thus, although the morphology does affect
frequency scaling, it does not account for the values observed
experimentally.

Another source of distortion in the frequency dependence of the V$_m$
is the fact that membrane permittivity (and capacitance) may also
depend on frequency~\cite{Cole-Cole1941,Hanai1965}.  Such a frequency
dependence is caused by a calorific dissipation during the
polarization of the membrane~\cite{Cole-Cole1941}, while the
Maxwell-Wagner phenomenon that we discuss here is a calorific
dissipation during the movement of charges on the membrane surface. 
However, direct capacitance measurements of bilipid membranes do not
evidence any significant variation of permittivity for frequencies
around 100~Hz~\cite{White1970}, and thus cannot explain the observed
deviations between cable theory and experiments shown in
Fig.~\ref{spectra}. Moreover, these
measurements~\cite{White1970,Alvarez1978} were realized on
artificially reconstructed membranes, which have a much simpler
structure compared to neuronal membranes (no saccharides, no
proteins, etc).  This is compatible with the possibility that in
biological membranes, the Maxwell-Wagner effect may be particularly
prominent.  The dependence of the membrane capacitance $c_m$ on
frequency may explain the flattening of the PSD above 1000~Hz, which
is visible in the experimental PSDs (see Fig.~\ref{comparaison}). 
However, the most likely explanation for this flattening is that the
recording is dominated by instrumental noise at such frequencies
(note that the bending of the experimental PSD above 4000~Hz in
Fig.~\ref{comparaison} is likely due to the low-pass 5~kHz filter
used during data acquisition).

Other factors may also affect the frequency scaling.  Taking into
account the finite rise time of synaptic events by using double
exponential templates amounts to add a factor 2 to the exponent
$\alpha$~\cite{Pow2003}. Similarly, introducing correlations in the
presynaptic activity may also affect the frequency scaling of V$_m$
power spectra~\cite{Marre2007}.  In all these cases, however, the
change in the scaling always consists in increasing the exponent
$\alpha$, while a decrease is needed to account for $\alpha$=2.5
scaling.

Thus, the frequency scaling of the V$_m$ activity can be affected by
several factors as discussed above.  Our results show that the
non-ideal character of the neuronal membrane can account for the
observed frequency scaling.  We believe that in reality, a
combination of factors is responsible for the observed frequency
scaling, and future experiments should be designed to test which are
the most determinant on frequency scaling, and what are the
consequences on the integrative properties of neuronal cable
structures.

Finally, our results show that the frequency-dependence of the
steady-state voltage profile (Fig.~\ref{atten}) is also affected by
the non-ideal character of the membrane capacitance.  Simulations
show that high-frequency signals ($>$ 100 Hz) propagate over larger
distances in the non-ideal cable model compared to the standard cable
model.  This theoretical result may be important to understand the
propagation of high-frequency events such as the ``ripples''
oscillations~\cite{Ylinen95,Grenier2001} across dendritic structures.

\

In conclusion, we provided here an extension to cable equations which
incorporates the non-ideal character of the membrane capacitance.  We
showed that this extension yields several detectable consequences on
neurons.  First, it affects basic cable properties such as the
voltage attenuation profile, especially at high frequencies.  Second,
it radically changes the frequency-scaling properties of voltage
power spectra.  The observed frequency scaling is within the range
predicted by the non-ideal cable model.  Fitting the model to
experiments provides an estimate of how ``non-ideal'' is the membrane
capacitance, and the significant values of $\tau_M$ found here
suggest that indeed, neuronal membranes may be far from being ideal
capacitors.


\clearpage
\appendix
\section*{Appendix 1: Frequency scaling in the standard cable model}

In this Appendix, we overview the frequency scaling characteristics
of the PSD of the V$_m$ for the ball-and-stick model using the
standard cable equations.

\subsection*{Dendritic current source located close to the soma}

We first consider the ball-and-stick model with an isolated current
source located in the dendrite close to the soma.  From expression
(\ref{B6+}) (see Appendix~2), we have:
 $$
  (Z_2\oplus Z_3)_{l\approx 0}\approx  \lim_{l\rightarrow0}
  (Z_2\oplus Z_3))=Z_3 ~ ,
 $$
and from expression (\ref{B6}), when the distance $l$ from the source
to the soma is small, the impedance of the distal part of the
dendrite is given by
$$
   Z_1\approx
   \frac{\lambda r_i}{\kappa_s}coth(\frac{\kappa_s
   l_d}{\lambda})] ~ ,
$$
where $l_d$ is the length of the dendrite.  From expression
(\ref{B0}), for small $l$, we have
$$
V_E = F_A i_S \approx (\frac{\lambda r_i}{\kappa_s}
\parallel Z_3)i_S ~ ,
$$
where $\frac{\lambda r_i}{\kappa_s}$ is the input impedance of a
finite dendritic branch.  Thus, from expression (\ref{B6++}), for
small $l$, we obtain
$$
F_T(l,\omega)  \approx \lim_{l \rightarrow 0}F_T(l,\omega) =1 ~ .
$$
Because $F_B \simeq 1$, the membrane potential at the center of the
soma is given by
\begin{equation}
    V_{soma} =(\frac{\lambda r_i}{\kappa_s}\parallel Z_3)i_S
    \label{B6+++}
\end{equation}
when the current source is located close to the soma.

Thus, for high frequencies ($>$ 100~Hz), the PSD of the somatic V$_m$
scales as $1/f^\alpha$ with $\alpha ~\in ~]3,4[ $ for a exponential
current source located close to the soma.  This result is similar to
single-compartment models~\cite{Pow2003}.

\subsection*{General case of dendritic current source}

We now consider the general case of a current source located at an
arbitrary position in the dendritic branch of the ball-and-stick
model.  We have necessarily $F_T \neq 1$, resulting in a
supplementary dependence on frequency.  Moreover, the current divider
$F_A$ also depends on frequency.  Numerical simulations show that the
PSD of the somatic V$_m$ scales as $1/f^\alpha$ with an exponent
$\alpha>3$.  For example, with exponential currents uniformly
distributed on a dendrite of $l_d$ = 500~$\mu$m, the frequency
scaling is close to an exponent of $\alpha$=4 (see continuous curve
in Fig.~\ref{PSDscaling}).  We verified numerically (not shown) that
the standard cable model cannot give a frequency scaling with a slope
smaller than $\alpha=3$ (using Poisson-distributed synaptic inputs).

A similar scaling with an exponent $\alpha=4$ was observed earlier,
when simulating realistic dendritic morphologies based on
reconstructed cortical pyramidal neurons~\cite{Pow2003}.

\section*{Appendix 2: Impedance analysis of the ball-and-stick model}

In this appendix, we derive the expressions needed to study the
frequency dependence of the ball-and-stick model
(Fig.~\ref{ball-stick}A), for both standard and non-ideal cable
equations.  The ball-and-stick model consists of a soma, which is
assumed to be the recording site, and a dendritic branch which
contains the source.  Referring to Fig.~\ref{ball-stick}A, we have
the source (S) and the recording locations (P), as well as the
impedances corresponding to the different regions ($Z_1$ for the
distal part of the dendrite, away of the source, $Z_2$ for the
proximal part of the dendrite, between the source and the soma, and
$Z_3$ for the soma).

We first evaluate the voltage at the current source:
\begin{equation}
V_s = i_s~\frac{Z_1(Z_2 \oplus Z_3)}{Z_1+(Z_2 \oplus
Z_3)}=F_A \ i_s ~, \label{B0}
\end{equation}
where the term $(Z_2 \oplus Z_3)$ is the input impedance of the
dendritic segment in series with $Z_3$. $F_A$ is the input impedance
as seen by the current source $i_s$ located at a position $l_s$ on
the dendritic branch.  Expression~\ref{B0} shows how $F_A$ varies as
a function of the position of the source in the dendrite.

Next, we calculate the somatic voltage from the transfer function
of the dendritic branch, $F_T$, which links the voltage at the source
with the somatic voltage.
\begin{equation}
    V_{soma}=F_T \ V_E
\end{equation}

Finally, we calculate the voltage transferred to the soma from the
equivalent circuit (Fig.~\ref{equiv}).
\begin{equation}
    V_P =\frac{Z_{3b}}{Z_{3a}+Z_{3b}} V_{soma}= F_B \ V_{soma} ~ ,
\end{equation}
where $F_B$ is the voltage divider caused by the fact that the tip of
the recording pipette is located inside the soma at some distance
from the membrane (in case of sharp-electrode recordings).  This
divider is entirely resistive and very close to 1, which expresses
the fact that the exact position of the pipette is not a determining
factor in the value of $V_P$.

\ \\ \centerline{------------------------ Figure~\ref{equiv} here
------------------------}

Thus, we have
\begin{equation}
V_{soma}=F_B \ F_T \ F_A \ i_s \simeq \ F_T \ F_A \ i_s
\end{equation}
We calculate these different terms below.

\subsection*{Input impedance $Z_1$ (distal part of the dendrite)}

For a current source $i_s$ located at position $l_s$, we have
\begin{equation}
i_s = i_{d_1}(l_s,\omega) + i_{d_2}(l_s,\omega) \label{B5.8-} ~ ,
\end{equation}
where $i_{d_1}(l_s,\omega)$ is the current density at the beginning
of the distal part of the dendrite (of length $\Delta l_1$)
, and $i_{d_2}(l_s,\omega)$ is the
current density of the proximal part of the dendrite (see
Fig(\ref{equiv}).  From expression~\ref{gen}, we have
$$ i_{d_1}(l_s,\omega) =
 -\frac{1}{r_i}\frac{\partial v_{m}(l_s +|\epsilon|,\omega)}{\partial x}
= \frac{\kappa}{\lambda r_i}(B -A) ~ ,
$$
where $|\epsilon|>0$ can be as small as desired.  This factor arises
because we consider point current sources, in which case the spatial
derivative of the $V_m$ is discontinuous at $x=l_s$.

 From the ``sealed end'' condition, we have
$$
i_{d_1}(l_s + \Delta l_1,\omega)= -\frac{1}{ r_i}\frac{\partial
v_{m}}{\partial x}(l_s+\Delta l_1,\omega) = -\frac{\kappa}{\lambda
r_i}~[A(\omega)~exp(-\frac{\kappa \Delta l_1}{\lambda})
-B(\omega)~exp(\frac{\kappa \Delta l_1}{\lambda})]=0 ~ .
$$
Thus, we have
\begin{equation}
A(\omega) =\frac{\lambda r_i}{\kappa}~\frac{exp(\frac{2\kappa
\Delta l_1}{\lambda})}{1-exp(\frac{2\kappa \Delta
l_1}{\lambda})}\label{B5.8}
\end{equation}
 and
\begin{equation}
B(\omega) =\frac{\lambda r_i}{\kappa}~\frac{1}{1-exp(\frac{2\kappa
\Delta l_1}{\lambda})} \label{B5.9}
\end{equation}
Consequently, we obtain
\begin{equation}
    Z_{1}=\frac{v_m(l_s,\omega)}{i_{d_1}(l_s,\omega)}
    =\frac{\lambda r_i}{\kappa}coth(\frac{\kappa \Delta l_1}{\lambda})
    \label{B6}
\end{equation}
where $\kappa = \kappa_s$ or $\kappa_{ext}$ for standard or non-ideal
cable models.

\subsection*{Input impedance $(Z_2\oplus Z_3)$ (proximal region)}

For the proximal part of the dendrite (of length $\Delta l_2 =
l_s$), which is in series with the impedance $Z_3$ at $x=0$ (see
Fig.~\ref{equiv}), we have (see expressions~\ref{B5.8-} and
\ref{gen})
$$ i_{d_2}(l_s,\omega) =
-\frac{1}{r_i}\frac{\partial v_{m}(l_s -|\epsilon|,\omega)}{\partial x}
= \frac{\kappa}{\lambda r_i}(B -A) ~ ,
$$
where $|\epsilon|>0$ can be as small as desired.

Moreover, we have
$$ i_{d_2}(0,\omega)= -\frac{1}{ r_i}\frac{\partial v_{m}}{\partial x}(0,\omega)
= -\frac{\kappa}{\lambda r_i}~[A(\omega)~exp(\frac{\kappa
l_s}{\lambda}) -B(\omega)~exp(-\frac{\kappa l_s}{\lambda})]=
\frac{v_m(0,\omega)}{Z_3}
$$
and
$$
    v_{m}(0,\omega) = A(\omega) ~exp(\frac{\kappa l_s}{\lambda})
     + B(\omega)~exp(-\frac{\kappa l_s}{\lambda})
$$
Thus, we obtain
$$
B(\omega) = A(\omega)~\frac{(1+\frac{\lambda r_i}{\kappa
Z_3})}{(1-\frac{\lambda r_i}{\kappa Z_3})}exp(\frac{2\kappa
l_s}{\lambda})
$$
and
$$
B(\omega) = A(\omega) + \frac{\lambda
r_i}{\kappa}i_{d_2}(l_s,\omega)
$$

Consequently, we obtain
\begin{equation}
A(\omega) =\frac{\lambda r_i~[ \kappa Z_3 -\lambda
r_i]i_{d_2}(l_s,\omega)}{\kappa ~[exp(\frac{2\kappa
l_s}{\lambda})+1][\kappa Z_3 ~tanh(\frac{\kappa
l_s}{\lambda})+\lambda r_i]}
\end{equation}
and
\begin{equation}
B(\omega) =\frac{\lambda r_i~[\kappa Z_3 +\lambda
r_i]i_{d_2}(l_s,\omega)~exp(\frac{2\kappa l_s}{\lambda})}{\kappa
~[exp(\frac{2\kappa l_s}{\lambda})+1][\kappa Z_3
~tanh(\frac{\kappa l_s}{\lambda})+\lambda r_i]} ~ .
\end{equation}
Thus, the input impedance $(Z_2\oplus Z_3)$ is given by:
\begin{equation}
(Z_2\oplus Z_3) = \frac{v_{m}(l_s,\omega)}{i_{d_2}(l_s,\omega)}
=\frac{\lambda r_i\cdot Z_3 }{[\kappa Z_3 ~tanh(\frac{\kappa
l_s}{\lambda})+\lambda r_i]} +
\frac{\lambda^2 r_i^{2}~tanh(\frac{\kappa l_s}{\lambda})~}{\kappa
~[\kappa Z_3 ~tanh(\frac{\kappa l_s}{\lambda})+\lambda r_i]}
\label{B6+} ~ ,
\end{equation}
where $\lambda=\sqrt{\frac{r_m}{r_i}}$ and $\kappa = \kappa_s ~
or~\kappa_{ext}$ according to which cable model is used.

For $Z_3 \to \infty$, we obtain the input impedance from
Eq.~\ref{B6}.

\subsection*{Calculation of the transfer function $F_T$}

To evaluate $F_T$, we calculate the voltage at point $x=l~$ by
imposing $v_m(l_s,\omega) =1$ at point $x=l_s$.  With this initial
value, the voltage $v_{m}(x)$ at point $x=0$ equals the value of the
transfer function at point $x=0$ (see Eq.~\ref{gen}).  In such
conditions, we obtain:
$$
 A(\omega) + B(\omega) =1~.
$$
Thus, we have
\begin{equation}
  F_T(x,\omega) = A(\omega) ~[exp(\frac{\kappa}{\lambda}(l_s-x)) -
  ~exp(-\frac{\kappa}{\lambda}(l_s-x))]+exp(-\frac{\kappa}{\lambda}(l_s-x))
\end{equation}
The voltage $v_m$ at point $x=0$ must equal
$Z_{3}~i_{i}(l,\omega)$ (current conservation).  We have
$$
\frac{\partial v_m}{\partial x}=-r_i ~i_i
$$
Consequently, we must obtain
\begin{equation}
   \frac{\partial F_T}{\partial x}\mid_{x=0}=-r_i ~\frac{v_m}{Z_{3}}\mid_{x=0}
   =\eta~v_m ~\mid_{x=0}=\eta~ F_T\mid_{x=0}
\end{equation}
where $\bf{\eta = -\frac{r_i }{Z_{3}}}$.  Thus, we have
\begin{equation}
    A(\omega) = \frac{(\kappa - \lambda\eta)exp(-\frac{\kappa l_s}{\lambda})}
    {\kappa[exp(\frac{\kappa l_s}{\lambda})+ exp(-\frac{\kappa l_s}{\lambda})]
     + \lambda\eta~ [exp(\frac{\kappa l_s}{\lambda}) - exp(-\frac{\kappa
     l_s}{\lambda})]}
\end{equation}
and the transfer function is given by
 \begin{equation}
  F_T(0,\omega) = A(\omega) ~[exp(\frac{\kappa}{\lambda} l_s) -
  ~exp(-\frac{\kappa}{\lambda} l_s)]+exp(-\frac{\kappa}{\lambda} l_s)
  \label{B6++}
 \end{equation}

Finally, we have
\begin{equation}
Z_{3}=Z_{3a}+\frac{R_m (i\omega C_m R_{sc}+1)}{i\omega C_m(
R_{sc}+R_m)+1}
\end{equation}
where $Z_{3a}$ is the plasma resistance in the soma.  $\kappa$
equals $\kappa_s$ or $\kappa_{ext}$ according to the cable model
considered.


\clearpage 


\clearpage

\begin{figure} 

\section*{Figure Legends}

\caption{Fall-off structure of power spectra of synaptic noise in
cortical neurons.  A. Time course of the membrane potential during
electrically-induced active states in a cortical neuron recorded
intracellularly from cat parietal cortex in vivo (data
from~\cite{Rud2005}).  B.  Power spectral density (PSD) of the
membrane potential in log scale.  The PSD has a fall-off structure
which follows a power law with a fractional exponent, around -2.6 in
this case (dashed line; modified from refs.~\cite{Dest2003,Rud2005}).
C. Four different morphologies of cortical pyramidal neurons from
cats obtained from previous studies~\cite{Contreras97,Douglas91}, and
which were incorporated into numerical simulations.  D. PSD obtained
from the four models in C, using the traditional cable formalism in
NEURON simulations.  The power-law exponent obtained was of 3.4, 3.3,
3.2 and 3.4, respectively (cells shown from left to right in C). }

\label{spectra}
\end{figure} 

\begin{figure} 

\caption{Different equivalent electric schemes for capacitors.  A.
Linear model of a capacitor, consisting of two resistances
($R_{sc}$ and $R_{pc}$), one inductance ($L_{sc}$) and one
capacitance element ($C$).  B. Approximation of the linear model
obtained by including a resistance ($R_{sc}$) in series with the
capacitance ($C$).  This leads to a characteristic relaxation time
for charging the capacitor (given by $\tau_M = R_{sc} C$).  C.
Ideal capacitance as in the standard cable model.}

\label{capacitor}
\end{figure} 

\begin{figure} 

  \caption{Comparison between $\kappa$ values in the standard and
non-ideal cable model.  The values of $\kappa$ are plotted for the
two models for various values of $\tau_M$ and two values of $\tau_m$
(5~ms and 20~ms).  The function $\kappa$ saturates for the non-ideal
cable model, and the value of the saturation equals to $\sqrt{1 +
\tau_m/\tau_M}$.  The $\kappa$ curves for the non-ideal model depart
from the standard model for a frequency that approaches the cut-off
frequency of $f_{c} = 1/\frac{1}{2\pi\tau_M}$.}

  \label{kappafig}
\end{figure} 

\begin{figure} 

  \caption{Steady-state voltage profile in a finite cable.  A cable
of 500~$\mu$m length and 2~$\mu$m diameter was considered with a
current source at $x=0$ (C$_m$ = 1~$\mu$F/cm$^2$; R$_i$ =
2~$\Omega$m). The voltage profiles in the non-ideal (gray lines) and
standard (black lines) cable models are compared for different
frequencies.  Two values of the membrane time constant are
considered, $\tau_m$ = 5~ms (A) and $\tau_m$ = 20~ms (B), which
correspond to two different conductance states ($\tau_M$ = 1.5~ms in
both cases, which corresponds to $\tau_M$ = 0.3~$\tau_m$ in A, and
$\tau_M$ = 0.075~$\tau_m$ in B).}

 \label{atten}
\end{figure}

\clearpage

\begin{figure} 

\caption{Ball-and-stick model used for calculations.  A. Scheme of
the ball-and-stick model where P indicates the soma, S the position
of the current source, and $Z_1$...$Z_3$ are impedances used in the
calculation. B. Example of a source current representing synaptic
bombardment in the ball-and-stick model. The current source consists
in Poisson-distributed exponential currents (see Materials and
Methods).  C. Power spectral density of the synaptic current source
shown in B. The PSD scales as a Lorentzian ($1/f^\alpha$ with an
exponent $\alpha=2$ between 100 and 400~Hz).}

\label{ball-stick}
\end{figure} 

\begin{figure} 

\caption{Power spectral density of the V$_m$ of the ball-and-stick
model with exponential synaptic currents uniformly distributed in the
dendrite (from 1 to 450~$\mu$~m, every 10~$\mu$~m).  The current
source of each synaptic event was the same and equals
$\exp(-t/0.1)$~nA, and the PSD is shown for the membrane potential at
the soma. The continuous curve shows the standard cable model, while
the other curves (dotted and dashed) show the non-ideal cable model
with different values of $\tau_M$.  Parameter values: C$_m$ =
1~$\mu$F/cm$^2$, $\tau_m$ = 5~ms, $l_d$ = 500~$\mu$m, R$_d$ =
1~$\mu$m, R$_{soma}$ = 7.5~$\mu$m,  R$_i$ = 2~$\Omega$m.}

\label{PSDscaling}
\end{figure} 

\begin{figure} 

\caption{Power spectral density of multiple synaptic events in the
ball-and-stick model. A. Voltage PSD at the some for a source current
similar to Fig.~\ref{ball-stick}B which was placed at different
positions in the dendrite (from top to bottom: 250 and 450~$\mu$m
from the soma).  For each location, the PSD is shown for the standard
cable model (gray) and for the non-ideal cable model (black).  B. PSD
obtained when the source currents were distributed in the dendrite
(from 1 to 450~$\mu$m, every 10~$\mu$m).  Parameter values: C$_m$ =
1~$\mu$F/cm$^2$, $\tau_m$ = 5~ms, $l_d$ = 500~$\mu$m, R$_d$ =
1~$\mu$m, R$_{soma}$ = 7.5~$\mu$m,  R$_i$ = 2~$\Omega$m, $\tau_M$ =
0.3 $\tau_m$.}

\label{sum}
\end{figure} 

\begin{figure} 

  \caption{Best fit of the non-ideal cable model to the power
spectral density obtained from intracellular experiments.  The
non-ideal cable model was simulated using a ball-and-stick model
subject to synaptic bombardment (see Materials and Methods).  The
dendritic branch had a 75~$\mu m$ length and the power spectral
density (PSD) was calculated from the somatic membrane potential.
Black: experimental PSD (see Fig.~\ref{spectra}); Gray: model PSD
(see Fig.~\ref{ball-stick}C for the PSD of the current source). The
slopes were calculated using a linear regression in the frequency
band 100--400~Hz.  The optimal value for $\tau_M$ was of $0.3
\tau_m$.  Parameter values: C$_m$ = 1~$\mu$F/cm$^2$, $\tau_m$ = 5~ms,
$l_d$ = 75~$\mu$m, R$_d$ = 1~$\mu$m, R$_{soma}$ = 7.5~$\mu$m,  R$_i$
= 2~$\Omega$m.}

  \label{comparaison}
\end{figure}

\begin{figure}

 \caption{Comparison of V$_m$ activities in the standard and
non-ideal cable models.  The current source is indicated on top,
while the bottom trace shows the V$_m$ activities superimposed.  The
inset shows a detail at 5 times higher temporal resolution.  Same
parameters as the optimal fit in Fig.~\ref{comparaison}.}

\label{Superpos}
\end{figure}

\begin{figure}

 \caption{Equivalent circuit for the ball-and-stick model.  $Z_1$ is
the input impedance of the dendritic branch (open circuit), $Z_2$ is
the impedance of the intermediate segment, in series with the
impedance $Z_3$ of the soma.}

\label{equiv}
\end{figure}


\clearpage

\begin{figure} 
  \centering
  \includegraphics[width=12cm]{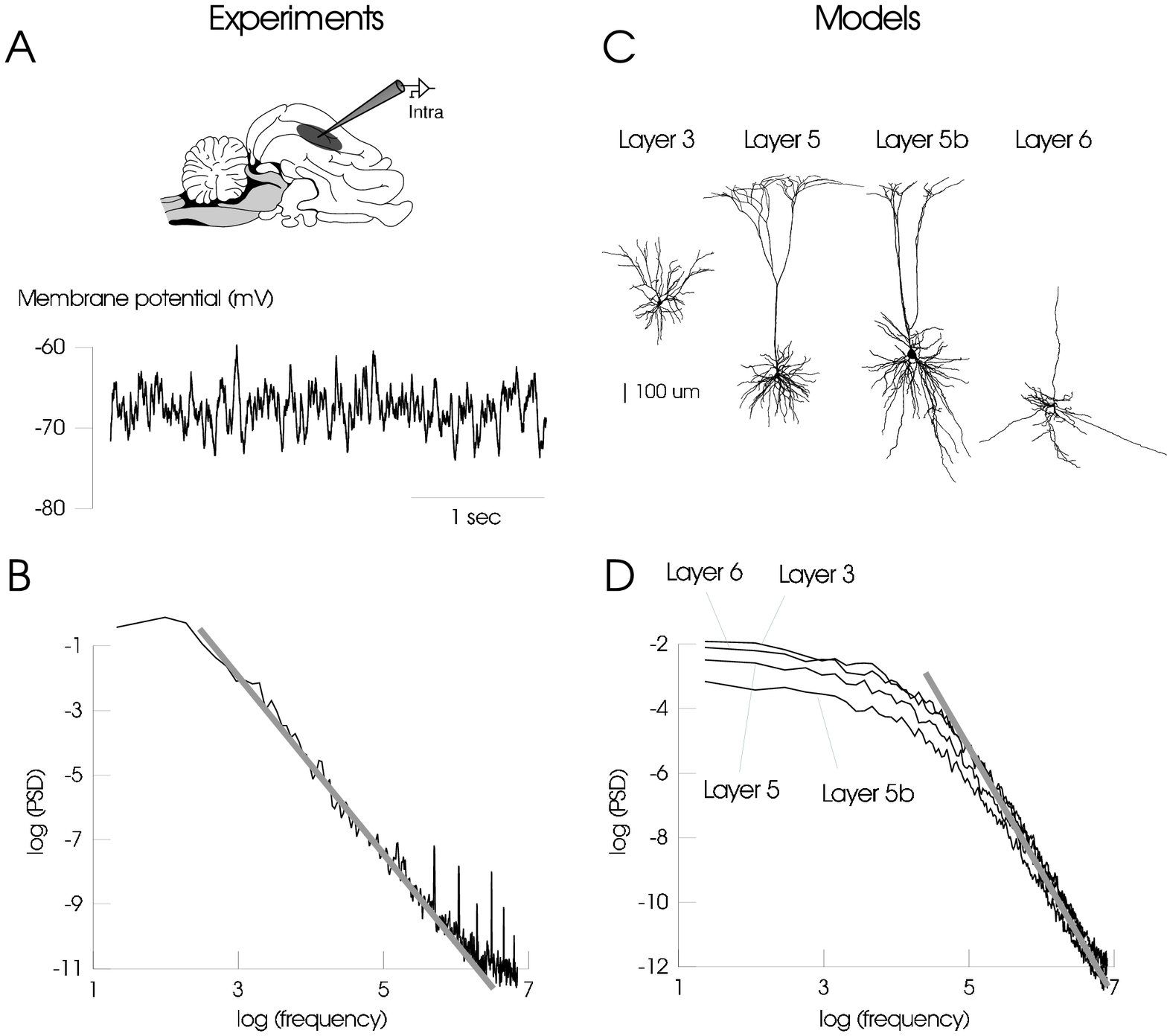}

\ \\ Figure~\ref{spectra}

\end{figure} 

\clearpage

\begin{figure} 
  \centering
  \includegraphics[width=10cm]{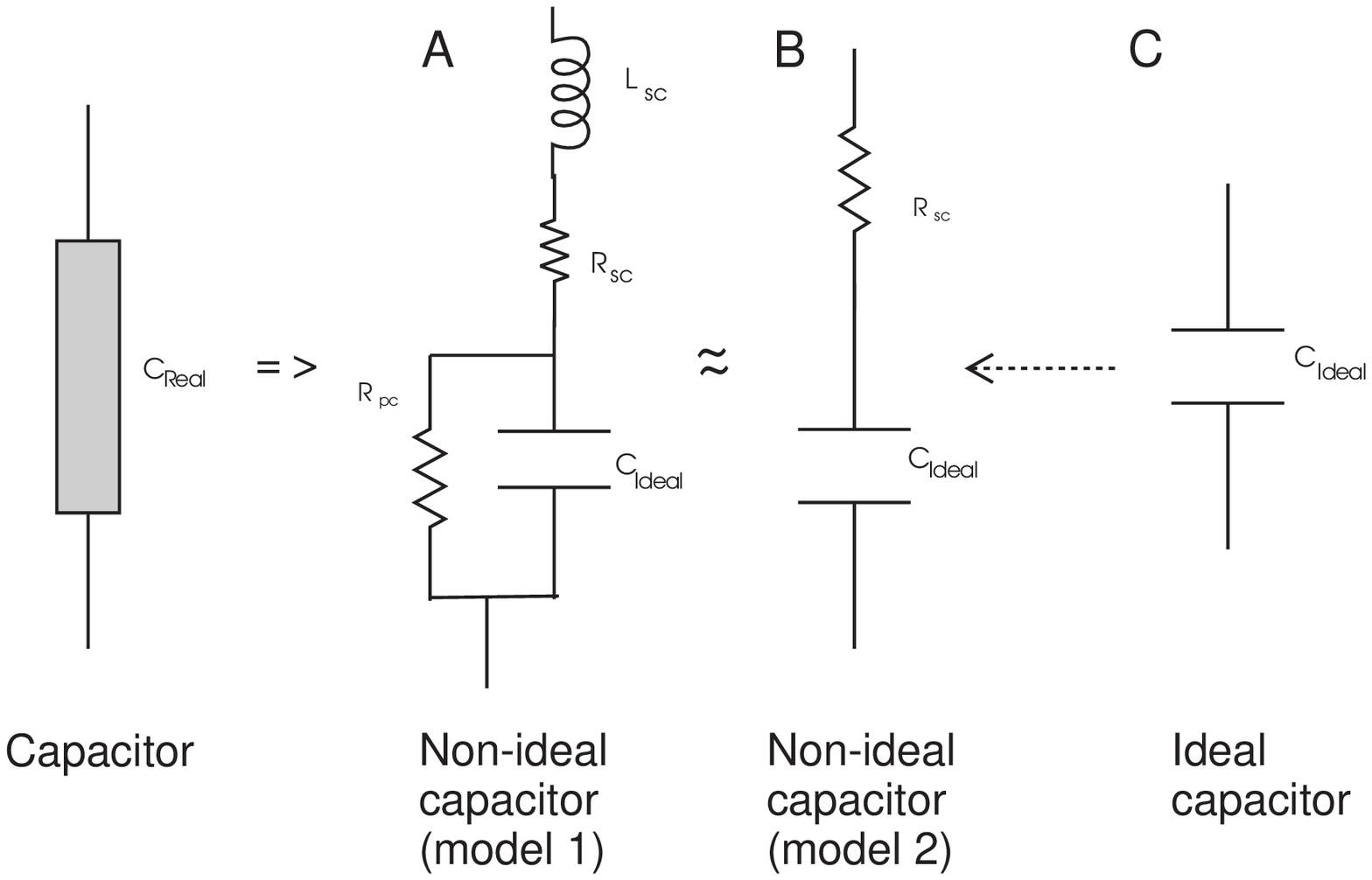}

\ \\ Figure~\ref{capacitor}

\end{figure} 

\begin{figure} 
  \centering
  \includegraphics[width=10cm]{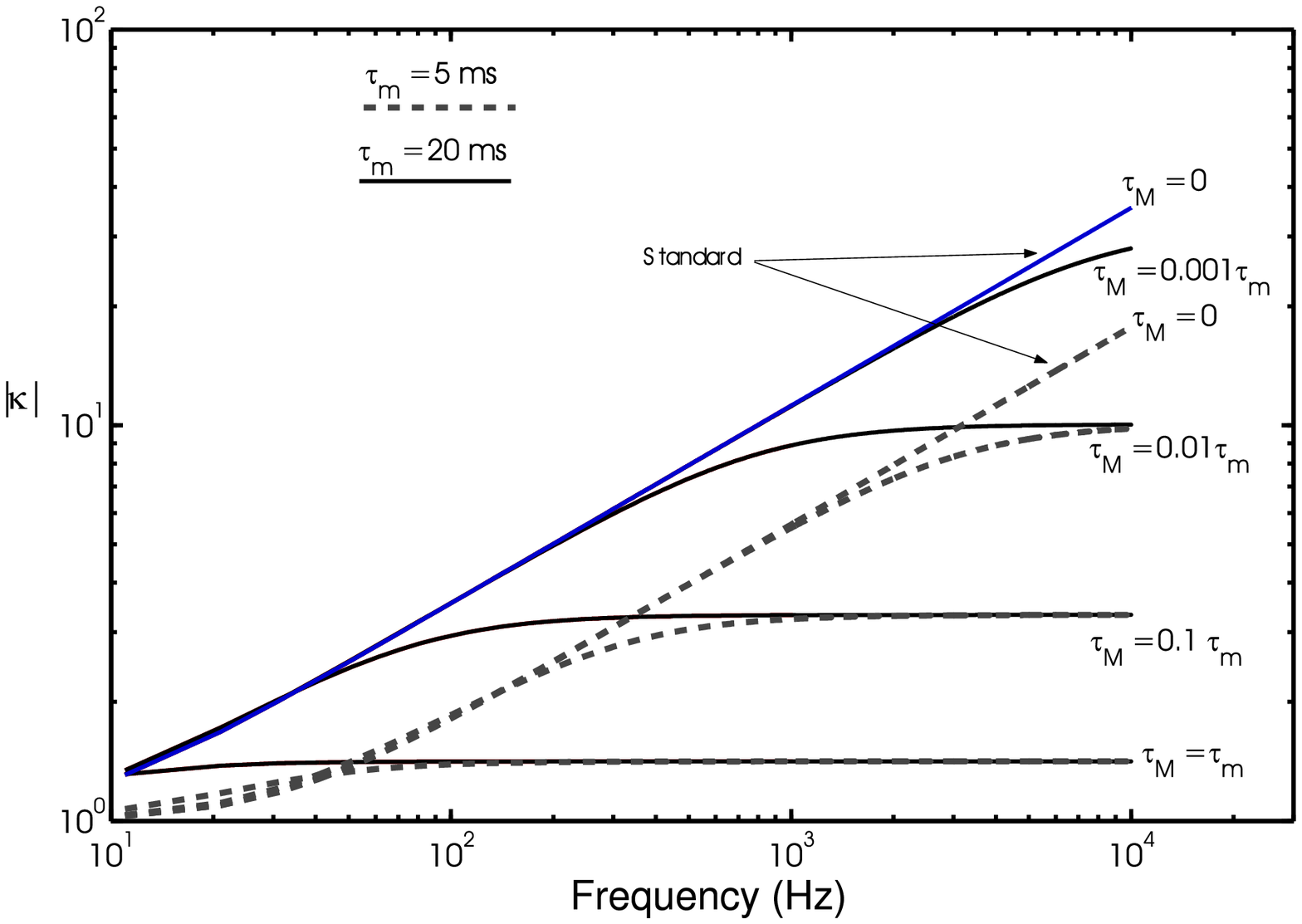}

\ \\ Figure~\ref{kappafig}

\end{figure} 

\begin{figure} 
\centering
  \includegraphics[width=8cm]{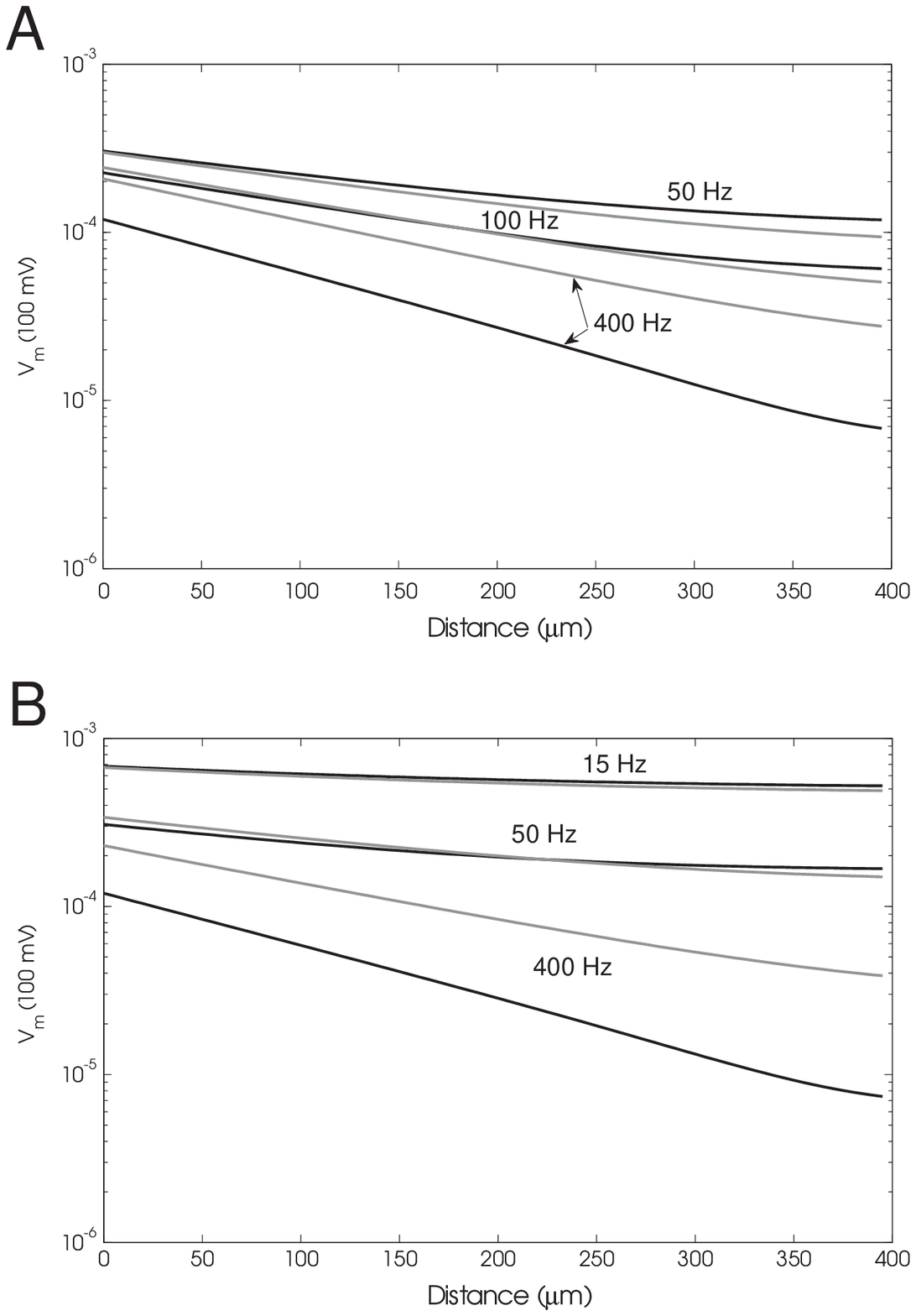}

\ \\ Figure~\ref{atten}

\end{figure}

\begin{figure} 
\centering
  \includegraphics[width=10cm]{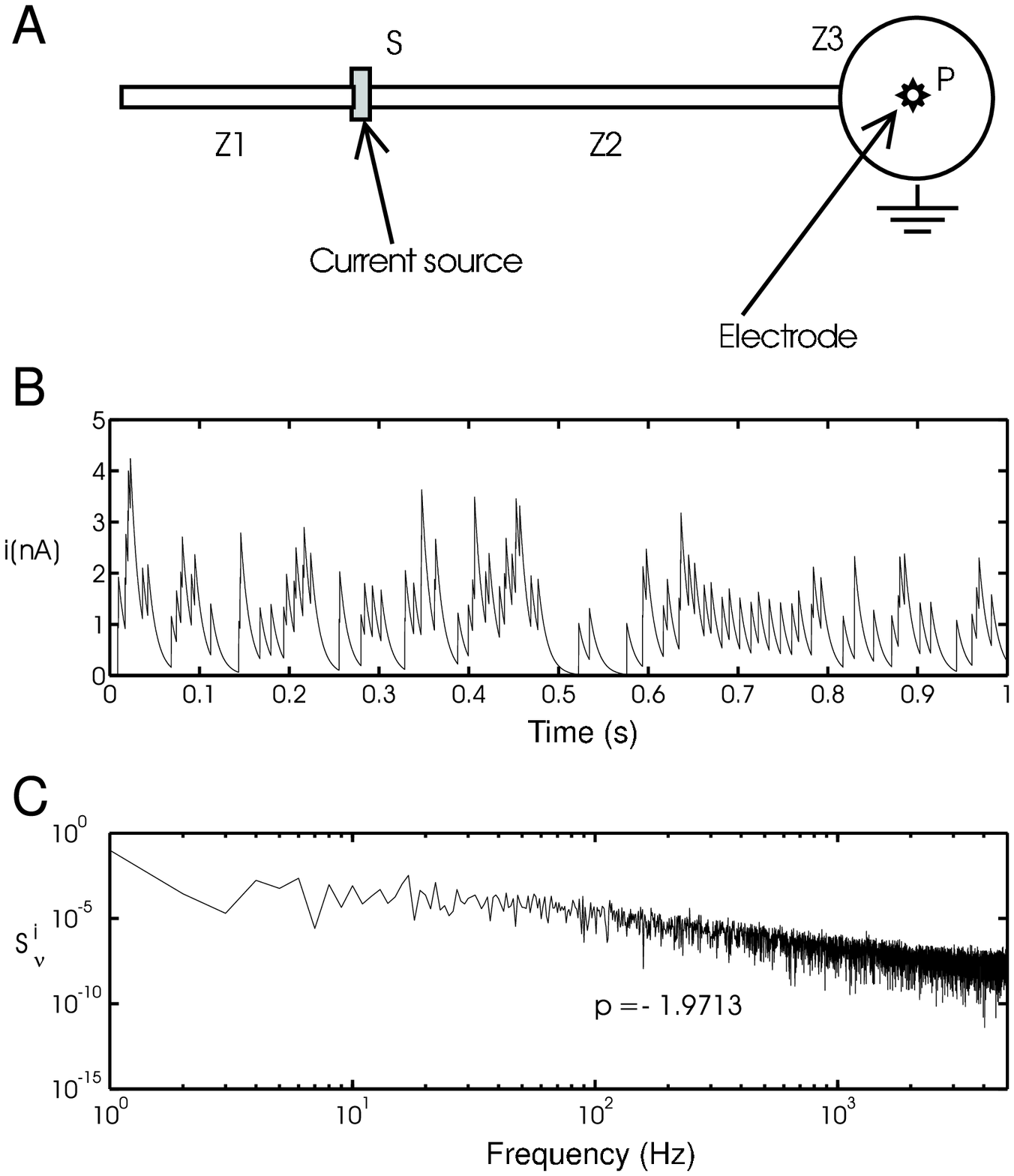}

\ \\ Figure~\ref{ball-stick}

\end{figure} 

\begin{figure} 
\centering
  \includegraphics[width=10cm]{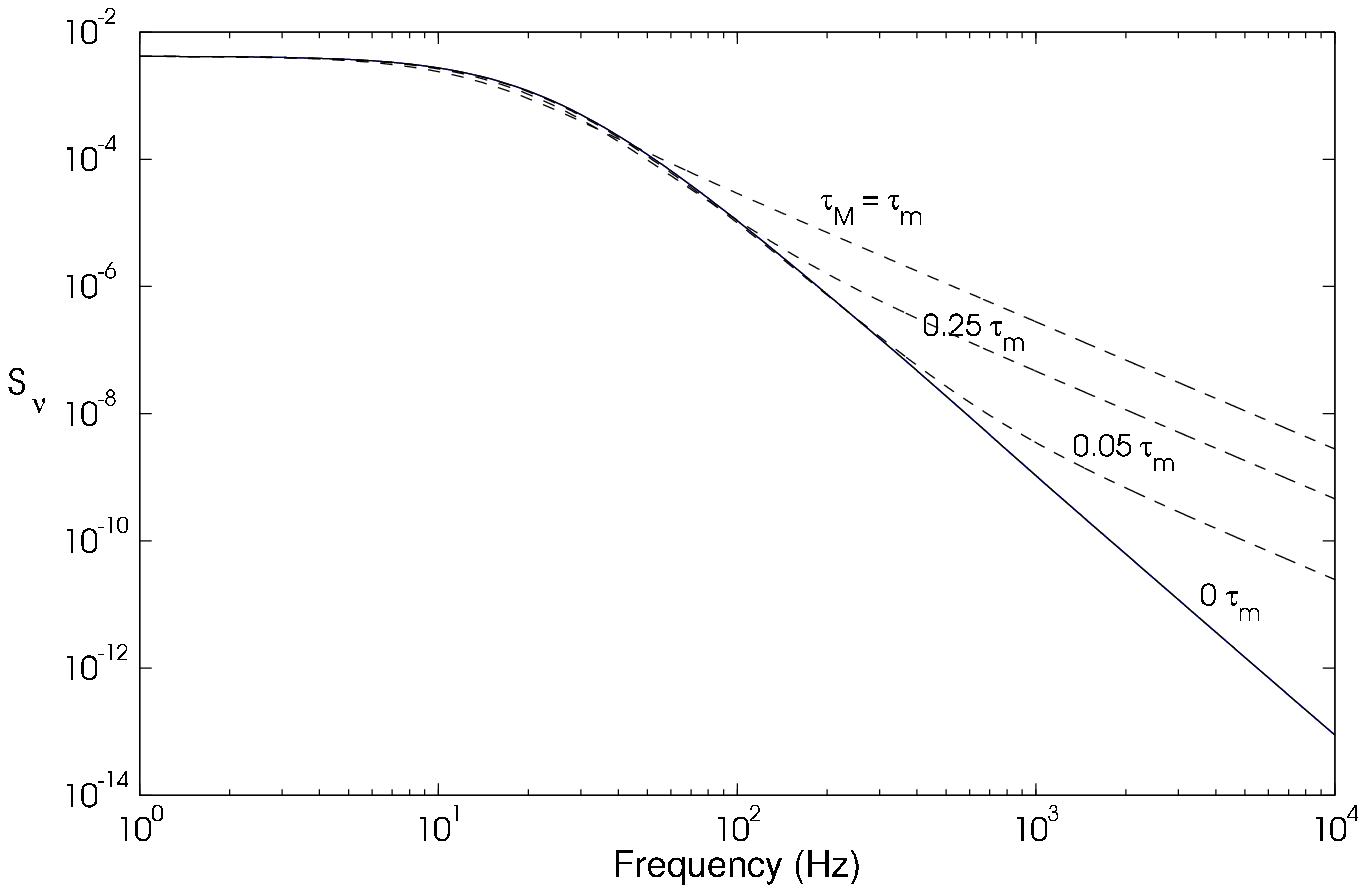}\\

\ \\ Figure~\ref{PSDscaling}

\end{figure} 

\begin{figure} 
\centering
  \includegraphics[width=8cm]{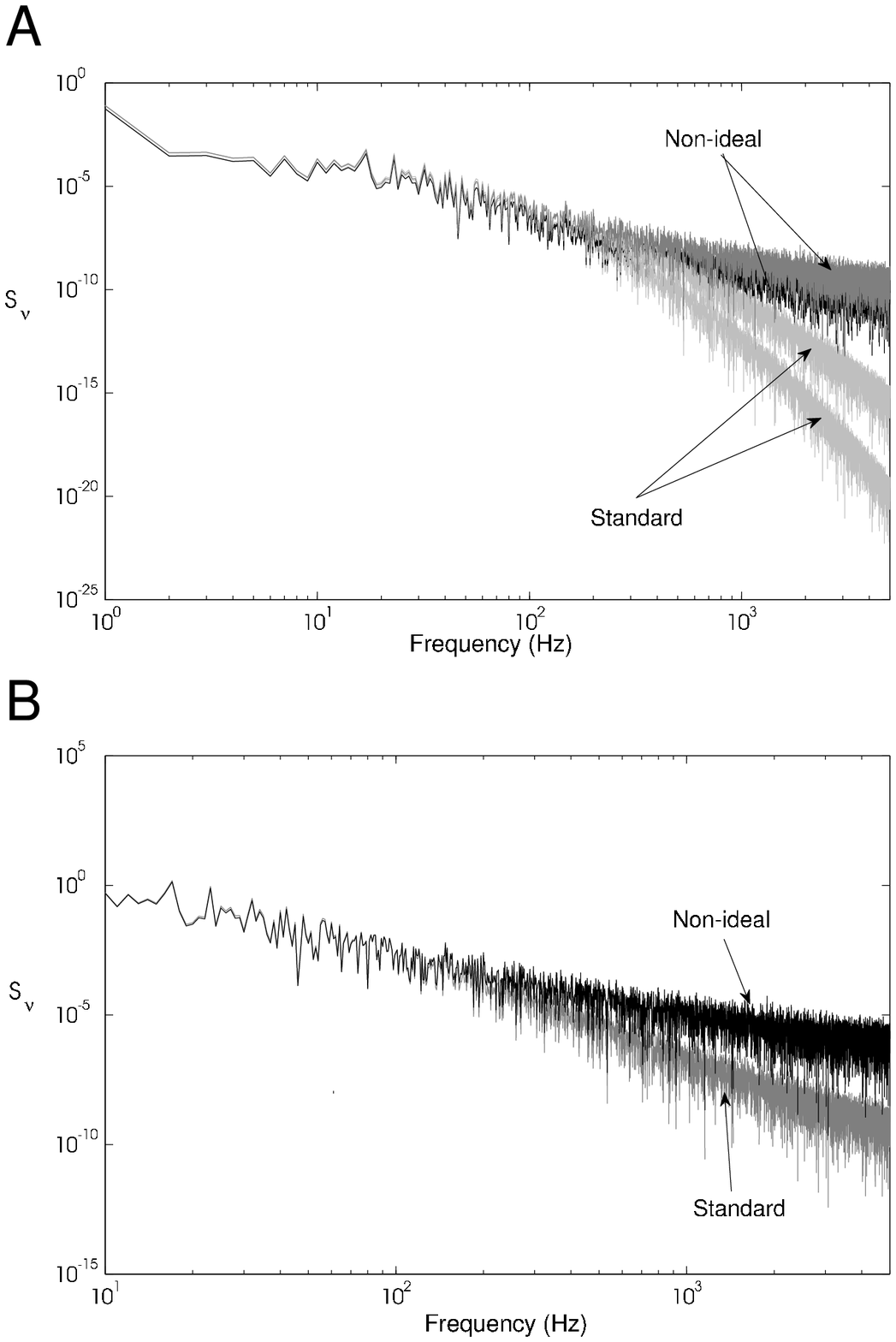}

\ \\ Figure~\ref{sum}

\end{figure} 

\begin{figure} 
\centering
\includegraphics[width=10cm]{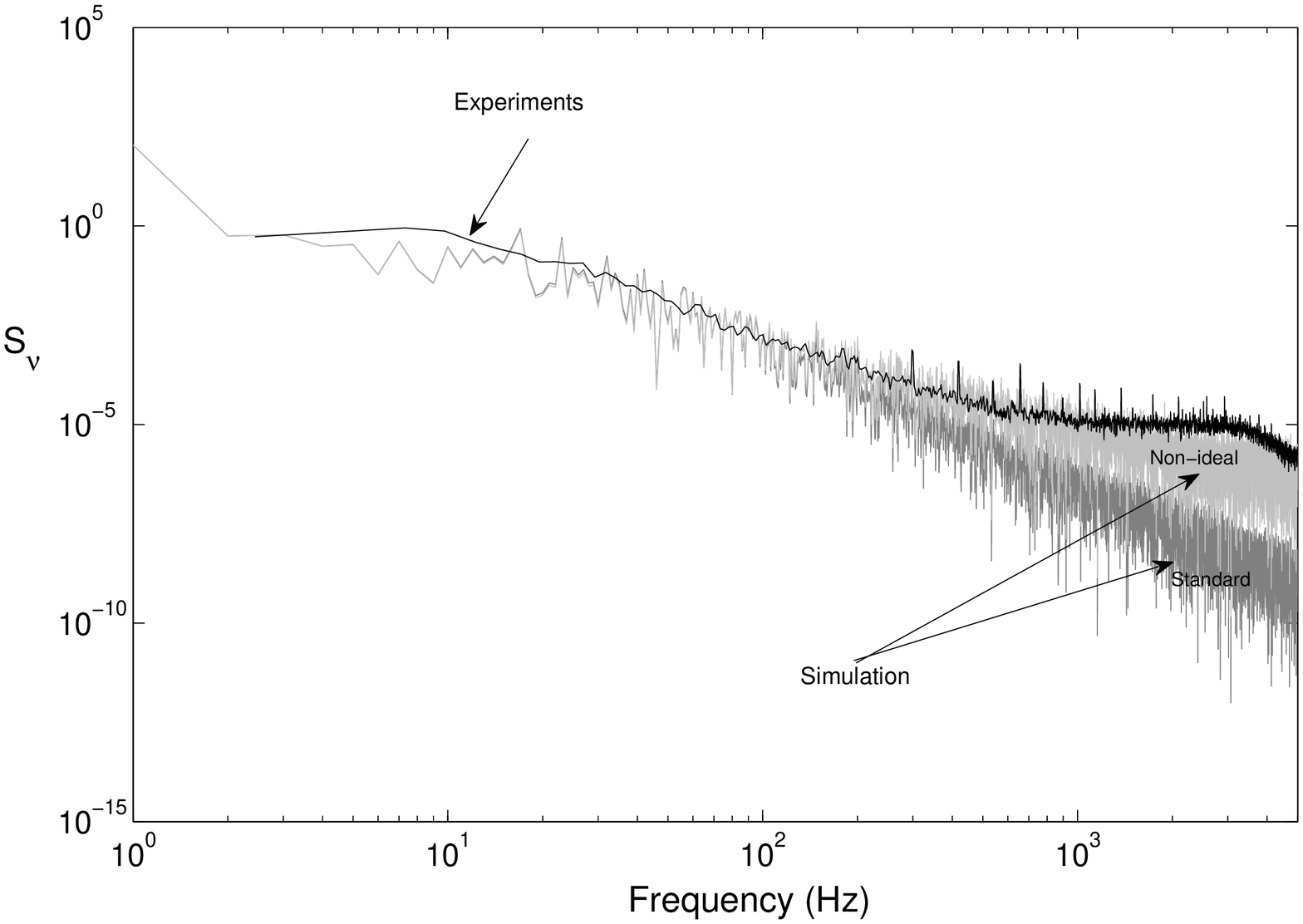}\\

\ \\ Figure~\ref{comparaison}

\end{figure}

\begin{figure} 
\centering
\includegraphics[width=12cm]{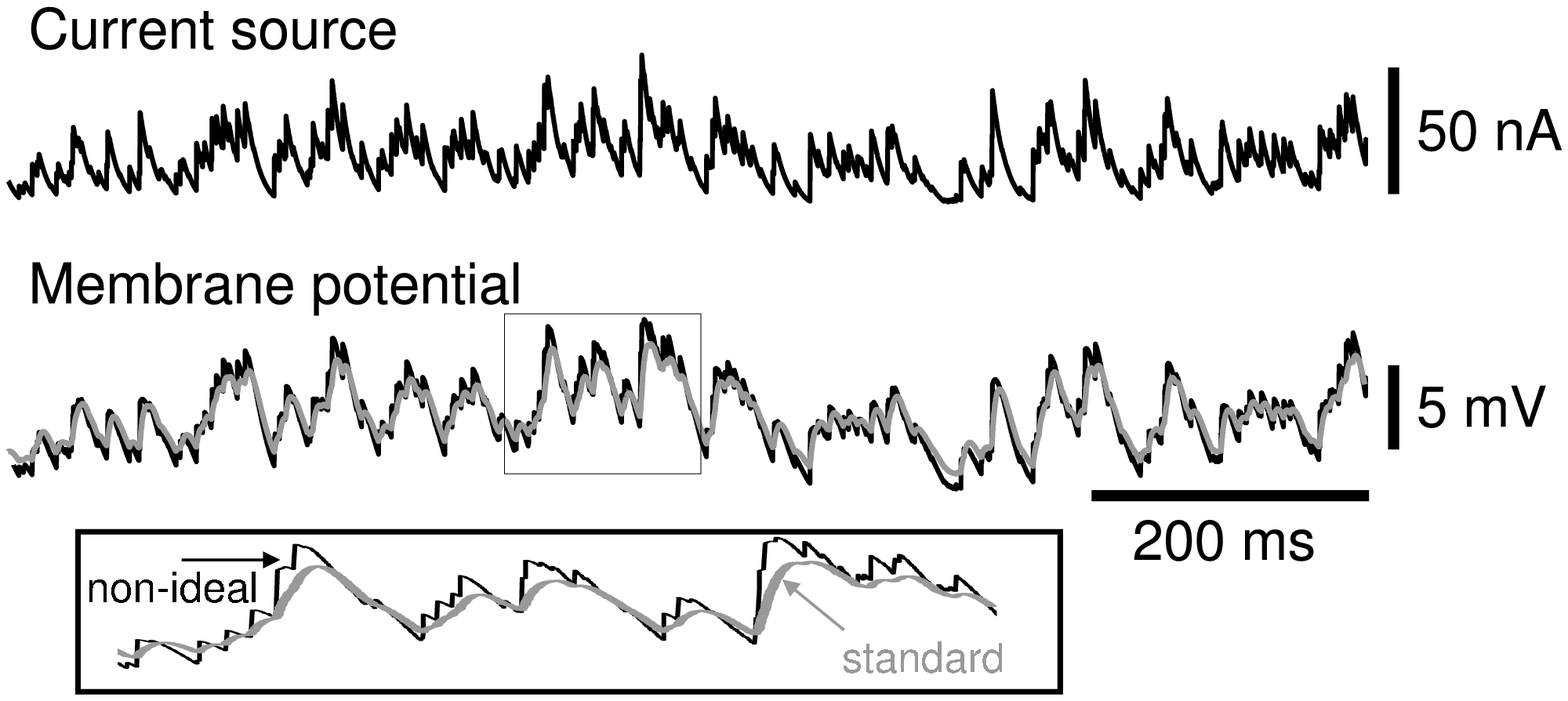}\\

\ \\ Figure~\ref{Superpos}

\end{figure}

\begin{figure}
 \centering
 \includegraphics[width=7cm]{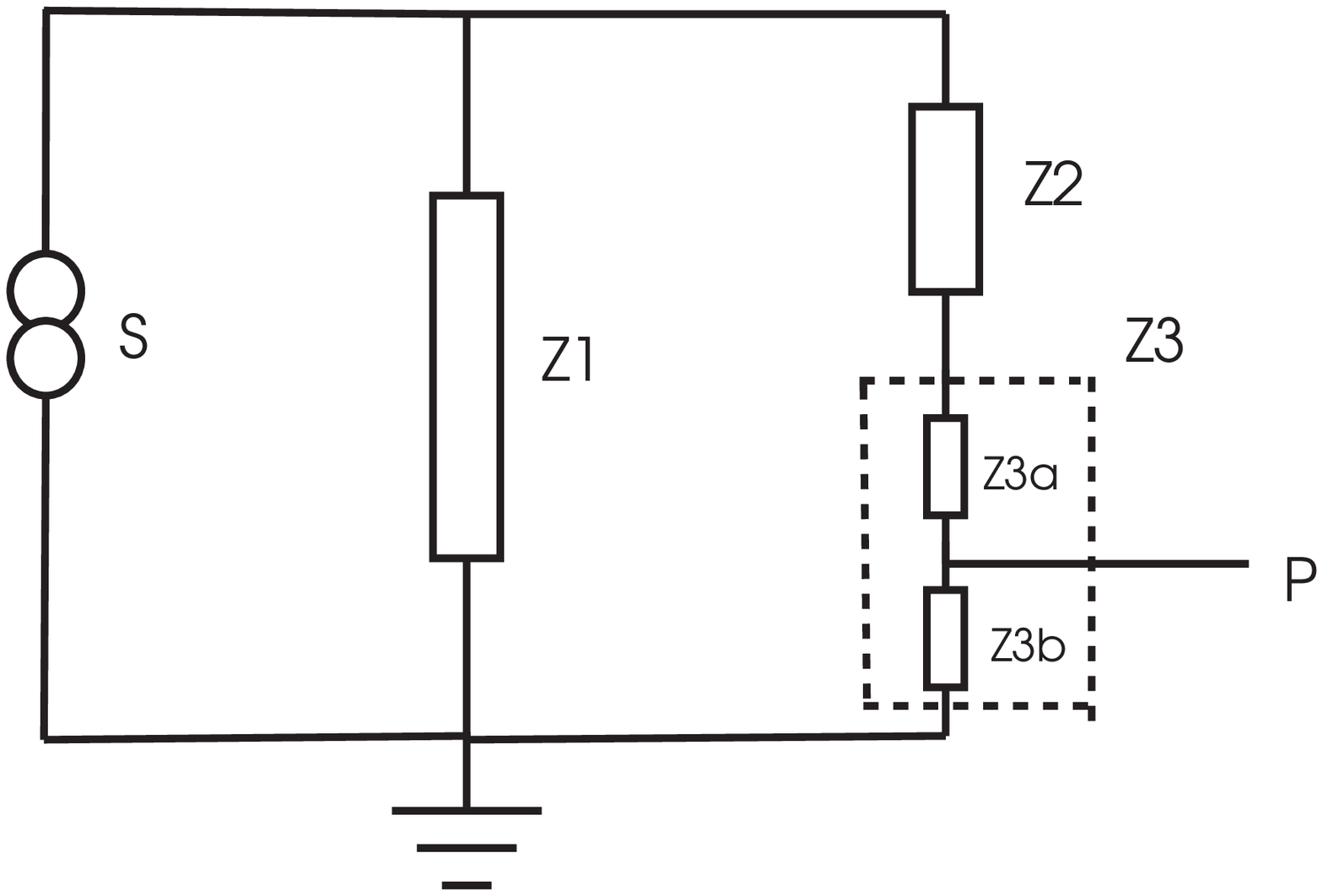}

\ \\ Figure~\ref{equiv}

\end{figure}


\begin{thebibliography}{99}

\small

\bibitem{Rall95} Rall, W. 1995. {\it The Theoretical Foundation of
Dendritic Function} (Segev, I., J. Rinzel and G.M. Shepherd, ed). MIT
Press, Cambridge MA.

\bibitem{Johnston95} Johnston, D. and S.M. Wu. 1995. {\it Foundations
of Cellular Neurophysiology}, MIT Press, Cambridge MA.

\bibitem{Rev2006} Brette, R., M. Rudolph, T. Carnevale, M. Hines, D.
Beeman, J.M. Bower, M. Diesmann, A. Morrison, P.H. Goodman, F.C.
Harris Jr., M. Zirpe, T. Natschlager, D. Pecevski, B. Ermentrout, M.
Djurfeldt, A. Lansner, O. Rochel, T. Vieville, E. Muller, A.
Davison, S. El Boustani and A. Destexhe.  2007.  Simulation of
networks of spiking neurons: A review of tools and strategies. {\it
J. Computational Neurosci.}, in press.  (article available at
\url{http://arxiv.org/abs/q-bio.NC/0611089}).

\bibitem{Dest2003} Destexhe A., M. Rudolph and D. Par\'e. 2003.  The
high-conductance state of neocortical neurons {\it in vivo}.  {\it
Nature Reviews Neurosci.} {\bf 4}: 739-751.

\bibitem{Diba2004} Diba, K., H.A. Lester and C. Koch. 2004. Intrinsic
noise in cultured hippocampal neurons: experiment and modeling. {\it
J. Neurosci.} {\bf 24}: 9723-9733.

\bibitem{Jacobson2005} Jacobson, G.A., K. Diba, A. 
Yaron-Jakoubovitch, C. Koch, I. Segev I and Y. Yarom. 2005. 
Subthreshold voltage noise of rat neocortical pyramidal neurones. 
{\it J. Physiol.} {\bf 564}: 145-160.

\bibitem{Rud2005} Rudolph M., J.G. Pelletier, D. Par\'e and A.
Destexhe. 2005. Characterization of synaptic conductances and
integrative properties during electrically-induced EEG-activated
states in neocortical neurons in vivo.  {\it J. Neurophysiol.} {\bf
94}: 2805-2821.

\bibitem{Pow2003} Destexhe, A. and M. Rudolph.  2004. Extracting
information from the power spectrum of synaptic noise.  {\it J.
Computational Neurosci.} {\bf 17}: 327-345.

\bibitem{Cole-Cole1941} Cole, K.S. and R.H. Cole. 1941.  Dispersion
and absorption in dielectrics. I. Alternating current
characteristics.  {\it J. Chem. Phys.} {\bf 9}: 341-351.

\bibitem{White1970} White, S.N. 1970.  A study of lipid bilayer
membrane stability using precise measurements of specific
capacitance.  {\it Biophys. J.} {\bf 10}: 1127-1148.

\bibitem{Eisenberg1980} Eisenberg, R.S. and R.T. Mathias.  1980. 
Structural analysis of electrical properties. {\it Crit. Reviews
Bioeng.} {\bf 4}: 203-232.

\bibitem{Alberts2002} Alberts, B, A Johnson, J Lewis, M Raff, K
Roberts and P Walter. 2002. {\it Molecular Biology of the Cell,
Fourth Edition.} Garland Publishing, New York.

\bibitem{SFN2007} Destexhe, A. and C. Bedard. 2007. A non-ideal
cable formalism which accounts for fractional power-law frequency
scaling of membrane potential activity of cortical neurons. {\it
Soc.  Neurosci. Abstracts} {\bf 33}: 251.14.

\bibitem{Contreras97} Contreras, D., A. Destexhe, and M.  Steriade.
1997.  Intracellular and computational characterization of the
intracortical inhibitory control of synchronized thalamic inputs {\it
in vivo}. {\it J.  Neurophysiol.} {\bf 78}: 335-350.

\bibitem{Douglas91} Douglas R.J., K.A. Martin and D.  Whitteridge. 
1991. An intracellular analysis of the visual responses of neurones
in cat visual cortex. {\it J. Physiol.} {\bf 440}: 659-696.

\bibitem{Hines97} Hines, M.L. and N.T. Carnevale. 1997.  The NEURON
simulation environment. {\it Neural Computation} {\bf 9}: 1179-1209.

\bibitem{Destexhe99} Destexhe, A. and D. Par\'e. 1999. Impact of
network activity on the integrative properties of neocortical
pyramidal neurons in vivo. {\it J.  Neurophysiol.} {\bf 81}:
1531-1547.

\bibitem{Bowick82} Bowick, C. 1982. {\it RF Circuit Design}, Newnes
Elsevier, New York.

\bibitem{Alvarez1978} Alvarez O. and R. Latorre. 1978.  t
Voltage-dependent capacitance in lipid bilayers made from monolayers.
{\it Biophys. J.} {\bf 21}: 1-17.

\bibitem{Raghuram1990} Raghuram, R. 1990. {\it Computer Simulation
of Electronic Circuits}, John Wiley \& Sons, New York.

\bibitem{Raju2003} Raju, G.G. 2003. {\it Dielectrics in Electric
Fields}, CRC Press, New York.

\bibitem{Bed2006} Bedard, C., A. Destexhe and H. Kroger. 2006. Model
of low-pass filtering of local field potentials in brain tissue. {\it
Physical Review E} {\bf 73}: 051911.

\bibitem{Hanai1965} Hanai T., D.A. Haydon and J. Taylor. 1965.  Some
further experiments on bimolecular lipid membranes.  {\it J.  Gen.
Physiol.} {\bf 48} (suppl): 59-63.

\bibitem{Marre2007} Marre, O., S. El Boustani, P. Baudot, M. Levy,
C.  Monier, N. Huguet, M. Pananceau, J. Fournier, A. Destexhe and Y.
Fr\'egnac. 2007.  Stimulus-dependency of spectral scaling laws in V1
synaptic activity as a read-out of the effective network topology. 
{\it Soc. Neurosci. Abstracts} {\bf 33}: 790.6.

\bibitem{Ylinen95} Ylinen, A., A. Bragin, Z. Nadasdy, G. Jando, I. 
Szabo, A. Sik and G. Buzsaki. 1995. Sharp wave-associated
high-frequency oscillation (200~Hz) in the intact hippocampus:
network and intracellular mechanisms. {\it J. Neurosci.} {\bf 15}:
30-46.

\bibitem{Grenier2001} Grenier, F., I. Timofeev and M. Steriade.
2001.  Focal synchronization of ripples (80-200~Hz) in neocortex and
their neuronal correlates.  {\it J. Neurophysiol} {\bf 86}:
1884-1898.

\end{thebibliography}
\end{document}